\documentstyle[12pt]{article}

\newcommand{\vp}{\varphi}
\newcommand{\ep}{\varepsilon}
\newcommand{\prt}{\partial}
\newcommand{\om}{\omega}
\newcommand{\Dlt}{\Delta}
\newcommand{\dlt}{\delta}
\newcommand{\ra}{\rightarrow}
\newcommand{\al}{\alpha}
\newcommand{\bt}{\beta}
\newcommand{\gm}{\gamma}

\begin{document}

\begin{center}
{\Large{\bf Topological Coherent Modes for Nonlinear Schr\"odinger 
Equation} \\ [5mm]
V.I. Yukalov$^1$ and E.P. Yukalova$^2$} \\ [3mm]
{\it $^1$ Bogolubov Laboratory of Theoretical Physics \\
Joint Institute for Nuclear Research, Dubna 141980, Russia \\ [2mm]
$^2$Laboratory of Informational Technologies \\
Joint Institute for Nuclear Research, Dubna 141980, Russia}

\end{center}

{\bf E-mail:} yukalov@thsun1.jinr.ru

\vskip 2cm

\begin{abstract}

Nonlinear Schr\"odinger equation, complemented by a confining 
potential, possesses a discrete set of stationary solutions. These 
are called coherent modes, since the nonlinear Schr\"odinger equation 
describes coherent states. Such modes are also named topological 
because the solutions corresponding to different spectral levels have 
principally different spatial dependences. The theory of resonant 
excitation of these topological coherent modes is presented. The 
method of multiscale averaging is employed in deriving the evolution 
equations for resonant guiding centers. A rigorous qualitative analysis 
for these nonlinear differential equations is given. Temporal behaviour 
of fractional populations is illustrated by numerical solutions.
\end{abstract}

\newpage

\section{Introduction}

Nonlinear Schr\"odinger equation has recently attracted a great deal 
of interest, since it provides an adequate description for collective
quantum states of trapped Bose atoms (see reviews [1--3]). This equation, 
as applied to Bose systems, is often termed the Gross-Pitaevskii
equation, who first considered that physical application [4,5].

For the purpose of accuracy in terminology, it is worth mentioning the 
following. As is possible to show [6], the nonlinear Schr\"odinger equation 
is an {\it exact equation for coherent states}. While the correct meaning of
the Gross-Pitaevskii equation [2] is that of an {\it approximate mean-field 
equation for the broken symmetry order parameter}.

Coherent states of trapped atoms are the solutions of the nonlinear 
Schr\"odinger equation with a confining potential. Stationary solutions
of this equation form a discrete set. Wave functions, corresponding to
different energy levels, are called [6] {\it coherent modes}. Different 
coherent modes possess qualitatively different spatial behaviour, because
of which they may be named {\it topological coherent modes}. Being the
solutions of the nonlinear Schr\"odinger equation, these modes have 
nothing to do with collective excitations described by the linear 
Bogolubov - De Gennes equations.

Bose-Einstein condensation of trapped atoms can be understood [3] as 
the condensation to the ground coherent state, that is, to the state 
with the lowest single-particle energy. In an equilibrium system, atoms 
always condense to the ground state. But, if an additional alternating 
field is switched on, with a frequency being in resonance with the 
transition frequency between two coherent energy levels, then higher 
topological modes can be excited, thus creating nonground-state 
condensates. The feasibility of such modes was first advanced in Ref. 
[7]. The alternating resonant field can be produced by modulating the 
confining potential of the trap or by imposing additional external 
fields. For instance, a rotating field for exciting vortices, can be 
created by multiple laser beams [8--10]. The resonant excitation can 
lead to the formation of several vortices [11] and, possibly, 
skyrmions in a spinor condensate [12]. The topological coherent modes 
have also been studied in Refs. [13,14]. Excitation of a dipole mode 
in a two-component condensate was observed [15]. The possibility of 
creating dark soliton states by means of the resonant excitation was
studied [11]. As is found, dark soliton states are unstable with respect
to their decomposition into several vortices. The states of multiple 
basic vortices are also more stable than a state of a single vortex with
a high winding number [3,16,17]. Investigating the temporal behaviour
[18,19] and collective excitations [20] of the formed coherent modes,
anomalous dynamical phenomena were found [18,19] reminding a kind of
critical phenomena. The origin of the latter has not been completely
understood.

The aim of the present paper is to develop the theory of the resonant 
formation of topological coherent modes and to provide an explanation 
of the dynamic critical phenomena. For this purpose, we give a rigorous 
stability analysis for the nonlinear evolution equations of guiding
centers. We demonstrate that dramatic changes in the dynamic properties
occur when crossing a saddle separatrix. The condition for the separatrix
touching the boundary defines a critical line on the parametric manifold.
This analysis is illustrated by the numerical solutions explicitly
displaying a dramatic qualitative change in the dynamics of fractional
populations when crossing the critical line.

\section{Resonant Excitation}

Consider a system of Bose atoms, which have experienced Bose-Einstein 
condensation under conditions typical of experiments with trapped atomic 
gases [1--3]. The coherent state of such atoms is described [6] by a
coherent field $\vp$ satisfying a nonlinear Schr\"odinger equation with 
the nonlinear Hamiltonian
\begin{equation}
\label{1}
\hat H(\vp) = -\; \frac{\hbar^2}{2m_0}\; {\bf\nabla}^2 + U({\bf r}) +
NA|\vp|^2 \; ,
\end{equation}
where $m_0$ is atomic mass, $U(\bf r)$ is a confining potential, $N$ is 
the number of trapped atoms, and $A\equiv 4\pi\hbar^2 a_s/m_0$ is the 
interaction strength, with $a_s$ being an $s$-wave scattering length.

Topological coherent modes are defined [7] as stationary solutions 
$\vp_n=\vp_n(\bf r)$ of the nonlinear Schr\"odinger equation
\begin{equation}
\label{2}
\hat H(\vp_n)\vp_n = E_n\vp_n \; .
\end{equation}
Here $n$ is a multi-index labelling the quantum states of coherent 
modes, and $E_n$ is a single-particle energy of the given coherent 
mode, which is normalized to unity as $(\vp_n,\vp_n)=1$. The excistence 
of the confining potential $U(\bf r)$ assumes that the spectrum $\{ E_n\}$
is discrete, hence the set $\{\vp_n\}$ of eigenfunctions is countable.

It is worth mentioning that the solutions to a nonlinear Schr\"odinger 
equation do not necessarily form a basis and, in general, are not orthogonal.
The property of being a basis has been proved for the eigenfunction sets
of only some one-dimensional nonlinear problems [21,22]. However, it is 
important to stress that we do not need and shall not use these properties
in what follows.

In order to induce intermode transitions, one has to impose an additional
time-dependent potential $\hat V =\hat V({\bf r},t)$ and to consider a
coherent field $\vp=\vp({\bf r},t)$ satisfying the temporal nonlinear
Schr\"odinger equation
\begin{equation}
\label{3}
i\hbar \; \frac{\prt\vp}{\prt t} =\left [ \hat H(\vp) + \hat V
\right ] \vp \; .
\end{equation}
Supposing that at the initial time all atoms were condensed to the 
ground-state coherent mode, one has the initial condition
\begin{equation}
\label{4}
\vp({\bf r},0) = \vp_0({\bf r}) \; .
\end{equation}
For transfering atoms from the ground state to a higher mode requires to 
impose an alternating potential 
\begin{equation}
\label{5}
\hat V = V_1({\bf r})\cos\om t + V_2({\bf r})\sin\om t \; ,
\end{equation}
with a frequency $\om$ being in resonance with the chosen transition 
frequency. If the wanted excited mode has the energy $E_k$, the transition
frequency is
\begin{equation}
\label{6}
\om_k \equiv \frac{1}{\hbar} \; \left ( E_k - E_0 \right ) \; .
\end{equation}
Then the resonance condition tells that the detuning of $\om$ from $\om_k$
is to be small,
\begin{equation}
\label{7}
\left | \frac{\Dlt\om}{\om}\right | \ll 1 \qquad (\Dlt\om \equiv 
\om - \om_k ) \; .
\end{equation}

One expects that, under the resonance condition (7), only the considered 
modes, connected by the resonance frequency (6), will be mainly involved 
in the process of excitation. This can be proved explicitly by invoking
the method of averaging [23]. For this purpose, let us look for the solution
of equation (3) in the form
\begin{equation}
\label{8}
\vp({\bf r},t) = \sum_n c_n(t) \vp_n({\bf r}) \exp\left ( -\; 
\frac{i}{\hbar} \; E_n t \right ) \; ,
\end{equation}
where $c_n(t)$ is a slowly varying amplitude, such that
\begin{equation}
\label{9}
\frac{\hbar}{E_n}\; \left | \frac{dc_n}{dt}\right | \ll 1 \; .
\end{equation}
The presentation (8) is to be substituted into equation (3), which is 
multiplied by $\exp(\frac{i}{\hbar}E_n t)$ and whose right-hand side is 
averaged over time. Integrating over time, the amplitudes $c_n$, according 
to condition (9), are treated as quasi-invariants. The integration of 
exponents yields 
$$
\lim_{T\ra\infty}\; \frac{1}{T}\; \int_0^T \exp\left\{ \frac{i}{\hbar}\;
(E_m - E_n ) t\right \}\; dt = \dlt_{mn} \; ,
$$
$$
\lim_{T\ra\infty}\; \frac{1}{T}\; \int_0^T \exp\left\{ \frac{i}{\hbar}\;
(E_m + E_k - E_n - E_l) t\right \}\; dt = \dlt_{mn}\dlt_{kl} +
\dlt_{ml}\dlt_{nk} - \dlt_{mk}\dlt_{kn}\dlt_{kl}   \; .
$$
This procedure results in the equation
\begin{equation}
\label{10}
\frac{dc_n}{dt} = - i \sum_{m(\neq n)} \al_{nm} | c_m|^2 c_n - \;
\frac{i}{2}\; \dlt_{n0}\bt_{0k} c_k e^{i\Dlt\om t} - \; \frac{i}{2}\; 
\dlt_{nk}\bt^*_{0k} c_0 e^{-i\Dlt\om t} \; ,
\end{equation}
in which the notation is used for the nonlinear transition amplitude
\begin{equation}
\label{11}
\al_{nm} \equiv A\; \frac{N}{\hbar} \; \int |\vp_n({\bf r})|^2 \left (
2|\vp_m({\bf r})|^2 - |\vp_n({\bf r})|^2 \right )\; d{\bf r}\; ,
\end{equation}
due to atomic interactions, and for the linear amplitude
\begin{equation}
\label{12}
\bt_{0k} \equiv \frac{1}{\hbar}\; \int \vp_0^*({\bf r}) \left [
V_1({\bf r}) - iV_2({\bf r})\right ] \vp_k({\bf r}) \; d{\bf r} \; ,
\end{equation}
caused by the resonant field (5). Note that the orthogonality of the modes
$\vp_n({\bf r})$ has nowhere been required. The initial condition to equation 
(10), in line with condition (4), is
\begin{equation}
\label{13}
c_n(0) = \dlt_{n0} \; .
\end{equation}

In the case when $n\neq 0,k$, the solution to equation (10) can be written as
$$
c_n(t) =c_n(0)\exp\left\{ - i\sum_{m(\neq n)} \al_{nm} 
\int_0^t |c_m(t')|^2\; dt'\right \} \; .
$$
This, in compliance with the initial condition (13), shows that $c_n(t)=0$
for all $n\neq 0,k$. So that only the behaviour of $c_0(t)$ and $c_k(t)$ is
nontrivial, with the related initial conditions 
\begin{equation}
\label{14}
c_0(0)=1 \; , \qquad c_k(0) = 0 \; .
\end{equation}
Thus, equation (10) reduces to the system of equations
$$
\frac{dc_0}{dt} = - i\al_{0k}|c_k|^2 c_0 -\; \frac{i}{2}\; \bt_{0k}
c_k e^{i\Dlt\om t} \; ,
$$
\begin{equation}
\label{15}
\frac{dc_k}{dt} = - i\al_{k0}|c_0|^2 c_k -\; \frac{i}{2}\; \bt_{0k}^*
c_0 e^{-i\Dlt\om t} \; ,
\end{equation}
with the initial conditions (14). The solutions to equations (15) are called
guiding centers.

\section{Change of Variables}

Equations (15) can be simplified by changing the variables. To this end, it is
convenient to introduce the population difference
\begin{equation}
\label{16}
s \equiv |c_k|^2 - |c_0|^2 \; .
\end{equation}
The amplitudes $c_0$ and $c_k$ can be presented as
$$
c_0 = \left ( \frac{1-s}{2}\right )^{1/2}\exp\left\{ i\left (\pi_0 +
\frac{\Dlt\om}{2}\; t\right )\right \} \; ,
$$
\begin{equation}
\label{17}
c_k = \left ( \frac{1+s}{2}\right )^{1/2}\exp\left\{ i\left (\pi_1 -\;
\frac{\Dlt\om}{2}\; t\right )\right \} \; ,
\end{equation}
with $\pi_0=\pi_0(t)$ and $\pi_1=\pi_1(t)$ being real phases.

Let us define the combination
\begin{equation}
\label{18}
\al\equiv \frac{1}{2}\; ( \al_{0k} + \al_{k0} )
\end{equation}
of the amplitudes (11), which is a real quantity, present the amplitude (12) 
as
\begin{equation}
\label{19}
\bt_{0k} \equiv \bt e^{i\gm} \qquad (\bt \equiv |\bt_{0k}|) \; ,
\end{equation}
and also define the renormalized detuning
\begin{equation}
\label{20}
\dlt \equiv \Dlt\om + \frac{1}{2}\; (\al_{0k} - \al_{k0} ) \; .
\end{equation}
Lastly, we introduce the phase variable
\begin{equation}
\label{21}
x \equiv \pi_1 -\pi_0 +\gm \; .
\end{equation}
The new functional variables (16) and (21) change in the rectangle
\begin{equation}
\label{22}
-1 \leq s \leq 1 \; , \qquad 0 \leq x \leq 2\pi \; .
\end{equation}
The related initial conditions are
\begin{equation}
\label{23}
s(0) = - 1 \; , \qquad x(0)=x_0 \; .
\end{equation}
The value $x_0=\pi_1(0)-\pi_0(0)+\gm$ can be any in the interval $[0,2\pi]$.
This is because, even if the initial phases of the considered modes were the 
same, the quantity $\gm$ can be varied by choosing an appropriate alternating 
potential (5). Also, even when $\pi_1(0)=\pi_0(0)$, the time dependence of
$\pi_1(t)$ and $\pi_0(t)$ is different, so that the evolution of $x=x(t)$
is not trivial.

With the new variables (16) and (21), equations (15) can be transformed to 
the Hamiltonian form characterized by the Hamiltonian
\begin{equation}
\label{24}
H(s,x) = \frac{\al}{2}\; s^2 - \bt\sqrt{1-s^2} \; \cos x  +\dlt s \; .
\end{equation}
The autonomous equations
$$
\frac{ds}{dt} = -\; \frac{\prt H}{\prt x} \; , \qquad 
\frac{dx}{dt} = \frac{\prt H}{\prt s}
$$
are identical to those that follow from equations (15) after substituting 
there expressions (17), and which are
\begin{equation}
\label{25}
\frac{ds}{dt} = -\bt\sqrt{1-s^2}\; \sin x \; , \qquad
\frac{dx}{dt} = \al s +\frac{\bt s}{\sqrt{1-s^2}}\; \cos x + \dlt \; .
\end{equation}
Equations (25) are more convenient for analyzing than those (15).

The autonomous Hamiltonian form of the evolution equations (25) tells us two
things. First, there is no dissipation for the coherent modes. The absence 
of intrinsic decoherence, connected with dissipation, here is quite clear.
For such a decoherence to arise, the Bose system is to be in contact with
an external bath [24,25] whose role can be played, e.g., by largely detuned
external laser beams [26] or by disturbing measurement instruments [27].
Decoherence of a coherent wave packet may appear owing to the 
number-of-particle fluctuations [28], which, actually, presupposes the 
existence of an external bath. The latter is a general cause of dissipation
for statistical systems [29,30] (other references on dissipative systems, can
be found in the recent review [31]).

Another conclusion which results from the Hamiltonian representation of the
autonomous evolution equations (25) is that in this dynamical system there 
can be no chaos. Hence the critical phenomena discovered [18,19] in the
dynamics of fractional populations cannot be attributed to the appearance of
chaotic motion. The origin of these critical phenomena will be elucidated in 
the following section.

\section{General Analysis}

To understand what happens with the solutions to the evolution equations (25)
under varying parameters, we have to analyze the general phase structure of 
these equations. For simplifying formulas, it is useful to introduce the
dimensionless parameters
\begin{equation}
\label{26}
b\equiv \frac{\bt}{\al} \; , \qquad \ep \equiv \frac{\dlt}{\al} \; .
\end{equation}
The first of the latter describes a relative intensity of the alternating
field (5), while the second parameter characterizes a relative value of the
detuning (20). These parameters can be positive as well as negative. The
sole thing is that the relative detuning will be treated as a small parameter,
$|\ep|\ll 1$.

Due to the existence of the integral of motion (24), the trajectory, starting
at the initial point (23) and defined by the equality $H(s,x)=H(s_0,x_0)$, 
writes
\begin{equation}
\label{27}
\frac{s^2}{2}\; - b\sqrt{1-s^2} \; \cos x + \ep s = \frac{1}{2}\; - \ep \; .
\end{equation}
The right-hand sides of equations (25), with the notation (26), can be
presented as
\begin{equation}
\label{28}
v_1 \equiv -b\sqrt{1-s^2}\; \sin x \; , \qquad
v_2 \equiv s + \frac{b s}{\sqrt{1-s^2}}\; \cos x + \ep \; .
\end{equation}
The stationary solutions to equations (25), given by $v_1=v_2=0$, are defined 
by the equations 
\begin{equation}
\label{29}
s^4 + 2\ep s^3 - (1 - b ^2 - \ep^2 ) s^2 - 2\ep s - \ep^2 = 0 \; ,
\qquad \sin x = 0 \; .
\end{equation}
The following analysis, keeping in mind the smallness of the detuning, will be
accomplished in the linear approximation with respect to $\ep$. Also, one has to
always remember that physically admissible fixed-point solutions are only those 
that are in the rectangle (22).

When $b^2\geq 1$, there exist the fixed points
$$
s_1^* = \frac{\ep}{b}\; , \qquad x_1^*=0 \; , 
$$
\begin{equation}
\label{30}
s_2^* = - \; \frac{\ep}{b}\; , \qquad x_1^*=\pi \; ,
\end{equation}
$$
s_3^* = s_1^*\; , \qquad x_3^*= 2\pi \; . 
$$

If $0\leq b<1$, the fixed points (30) continue to exist, but there arise two 
new points
$$
s_4^* = \sqrt{1 - b^2} + \frac{b^2\ep}{1-b^2} \; , \qquad x_4^* = \pi \; ,
$$
\begin{equation}
\label{31}
s_5^* = - \sqrt{1 - b^2} + \frac{b^2\ep}{1-b^2} \; , \qquad x_4^* = \pi \; .
\end{equation}

And if $-1< b\leq 0$, then the fixed points (30) again exist, but there appear
the additional points
$$
s_4^* =\sqrt{1 - b^2} + \frac{b^2\ep}{1-b^2} \; , \qquad x_4^*= 0 \; ,
$$
$$
s_5^* = -\sqrt{1 - b^2} + \frac{b^2\ep}{1-b^2} \; , \qquad x_5^*= 0 \; ,
$$
$$
s_6^* = s_4^* \; , \qquad x_6^*= 2\pi \; ,
$$
\begin{equation}
\label{32}
s_7^* = s_5^* \; , \qquad x_7^*= 2\pi \; .
\end{equation}
In this way, the value $b^2=1$ corresponds to a dynamical phase transition, 
when the structure of the phase portrait changes qualitatively.

To analyse the stability of motion, we consider the Jacobian expansion matrix
$\hat X = [X_{ij}]$ with the elements
$$
X_{11} \equiv \frac{\prt v_1}{\prt s} = \frac{b s}{\sqrt{1-s^2}}\; \sin x \; ,
\qquad X_{12} \equiv \frac{\prt v_1}{\prt x} = - b\sqrt{1 - s^2}\; \cos x \; ,
$$
$$
X_{21} \equiv \frac{\prt v_2}{\prt s} = 1+ \frac{b \cos x}{(1-s^2)^{3/2}}\; ,
\qquad X_{22} \equiv \frac{\prt v_2}{\prt x} = - X_{11} \; .
$$
The local expansion rate [32], defined as
$$
\Lambda(t) \equiv \frac{1}{t}\; {\rm Re} \int_0^t 
{\rm Tr}\; \hat X(t')\; dt' \; ,
$$
nullifies, $\Lambda(t)=0$, as it should be for Hamiltonian systems whose 
phase volume is conserved. The eigenvalues of the expansion matrix $\hat X$ 
are given by the equation
\begin{equation}
\label{33}
X^2 = \frac{b^2}{1-s^2} \left ( s^2 \sin^2 x - \cos^2 x\right ) -
b\sqrt{1-s^2}\; \cos x \; .
\end{equation}
Evaluating the eigenvalues at the fixed points, we shall employ expressions 
(30) to (32), limiting ourselves by the linear in $\ep$ approximation.

In the case of $b^2>1$, we have
\begin{equation}
\label{34}
X_1^\pm = \pm i\sqrt{b(b+1)} = X_3^\pm \; , \qquad 
X_2^\pm = \pm i\sqrt{b(b-1)} \; ,
\end{equation}
so that all fixed points (30) are centers.

When $0\leq b<1$, the first and third fixed points remain the centers, while 
the second fixed point $(s_2^*,x_2^*)$ becomes a saddle. The fixed points 
(31) are also the centers. The related exponents are
$$
X_1^\pm = \pm i\sqrt{b(1+b)} = X_3^\pm \; , \qquad 
X_2^\pm = \pm \sqrt{b(1-b)} \; ,
$$
\begin{equation}
\label{35}
X_4^\pm = \pm i\sqrt{1-b^2}\left [ 1 + \frac{(2+b^2)\ep}{(1-b^2)^{3/2}}
\right ]^{1/2} \; , \qquad
X_5^\pm = \pm i\sqrt{1-b^2}\left [ 1 - \frac{(2+b^2)\ep}{(1-b^2)^{3/2}}
\right ]^{1/2} \; .
\end{equation}
The saddle separatrices which are the trajectories that pass through the 
saddles and separate the phase regions with qualitatively different dynamic 
properties, are given by the equation $H(s,x)=H(s_2^*,x_2^*)$, which results 
in the separatrix equation
\begin{equation}
\label{36}
\frac{s^2}{2} \; - b\sqrt{1-s^2}\; \cos x + \ep s =  b \; ,
\end{equation}
defining two separatrices, since this is a square equation with respect to $s$.
The separatrix extremal points can be found from the equation
$$
\frac{ds}{dx} = -\; \frac{b(1-s^2)\sin x}{(s+\ep)\sqrt{1-s^2}+ bs\cos x}
= 0\; .
$$
As follows from the above equations, the lower separatrix parts touch the 
boundary at the points $s=-1$, $x=0,2\pi$, provided that
\begin{equation}
\label{37}
b+\ep = \frac{1}{2} \; .
\end{equation}

In this way, accepting as initial conditions $s_0=-1,\; x_0=0$, one has the 
following picture. For $b<0.5-\ep$, the motion is limited from above by the 
lower separatrix parts and from below, by the boundary $s=-1$. When the 
pumping parameter $b>0.5-\ep$, the trajectory passes to the phase region 
limited from above by the upper separatrix parts and from below by the lower 
separatrix parts. If $b=0.5-\ep$, the dynamical system is structurally 
unstable with respect to small variations of initial conditions. On the
manifold of the system parameters, the line (37) plays the role of a 
{\it critical line}. In the vicinity of this line, solutions to the evolution
equations display dramatic effects, when a tiny variation of a parameter
qualitatively changes the properties of solutions. This makes it possible to
call such dynamical effects the {\it critical dynamic phenomena} [33].

For the case when $-1<b\leq 0$, the expansion exponents, corresponding to the
fixed points (30) and (32), are
$$
X_1^\pm = \pm\sqrt{|b|(1+b)} = X_3^\pm \; , \qquad
X_2^\pm = \pm i \sqrt{|b|(1-b)} \; ,
$$
$$
X_4^\pm = \pm i\sqrt{1-b^2}\left [ 1 + 
\frac{(2+b^2)\ep}{(1-b^2)^{3/2}}\right ]^{1/2} = X_6^\pm \; ,  
$$
\begin{equation}
\label{38}
X_5^\pm = \pm i\sqrt{1-b^2}\left [ 1 - \; 
\frac{(2+b^2)\ep}{(1-b^2)^{3/2}}\right ]^{1/2} = X_7^\pm \; .
\end{equation}
Hence, the first and third fixed points become the saddles, while all other
points are the centers. The separatrices connecting the saddles are defined 
by the equations $H(s,x)=H(s_1^*,x_1^*)=H(s_3^*,x_3^*)$, which yield
\begin{equation}
\label{39}
\frac{s^2}{2}\; - b\sqrt{1-s^2}\; \cos x +\ep s = - b \; .
\end{equation}
The lower separatrix part touches the boundary at the point $s=-1$, $x=\pi$
under the condition
\begin{equation}
\label{40}
|b| +\ep =\frac{1}{2} \; .
\end{equation}
The phase picture, as compared to the previous case $0\leq b<1$, looks similar, 
but being shifted by $\pi$ along the axis $x$. Now, if the initial point would
be $s_0=-1$, $x_0=0$, the motion would be always limited from above by the 
lower separatrix parts and from below, by the boundary $s=-1$. No dramatic 
changes would happen when varying $b$. However, if the initial point is 
taken as $s_0=-1$, $x_0=\pi$, one again encounters the same critical 
dynamic phenomena on the critical line (40).

This analysis explains that the occurrence of critical dynamic phenomena is
caused by the existence of a critical line on the manifold of system parameters
and happens only under a special choice of initial conditions, when the latter
are touched by a separatrix. The initial conditions for the variable (21) can 
be varied in a wide diapason by choosing the appropriate alternating field 
(5), which would yield the related linear amplitude (12), with the required 
value of
$$
\gm = {\rm arg}\; \bt_{0k} = {\rm tan}^{-1}\; 
\frac{{\rm Im}\; \bt_{0k}}{{\rm Re}\; \bt_{0k}} \; ,
$$
defined in equation (19).

\section{Numerical Solution}

In order to explicitly illustrate the critical dynamic phenomena, occurring 
when crossing the critical line on the parametric manifold, we have 
accomplished numerical calculations of the fractional populations
$$
n_0(t)\equiv |c_0(t)|^2 = \frac{1-s(t)}{2} \; , \qquad
n_k(t)\equiv |c_k(t)|^2 = \frac{1+s(t)}{2} \; .
$$
This can be done by numerically solving either equations (15) or (25), which 
are equivalent. The results are presented in figures 1 to 3, where time is
measured in units of $\al^{-1}$, the pumping parameter $b=0.5$ is fixed, while
the detuning $\ep$ is varied so that to cross the critical line (37). The
initial conditions for the evolution equations are taken as $s_0=-1$, $x_0=0$,
that is $n_0(0)=1$, $n_k(0)=0$. 

In figure 1, the detuning is negative, so that one is slightly below the 
critical line (37). The fractional populations oscillate displaying a kind
of Rabi oscillations, if one looks for analogies with optics [34]. However,
since the problem considered here is nonlinear, there is no a well defined 
constant Rabi frequency, whose analog would be now a function of time [18,19]. 
In some sense, these oscillations could be named {\it nonlinear Rabi 
oscillations}. Approaching the critical line (37) by increasing the detuning 
$\ep$, the oscillation amplitude increases. The motion of the population
$n_k$ of the excited topological mode is limited from above by the lower parts 
of separatrices and from below, by the boundary $n_k=0$.

Figure 2 demonstrates the motion in a tiny vicinity of the critical line (37).
Changing the detuning from $\ep=10^{-4}$ to $\ep=0$ drastically transforms
the shape of oscillations. The oscillation period more than doubles and wide 
flat platos in the time dependence of the population appear. Crossing the
critical line, with a microscopic variation of the detuning from $\ep=0$ to 
$\ep=10^{-9}$, again yields a drastic transformation of the population shapes.
The period is again almost doubled; the upward casps of $n_k(t)$ and down-ward
casps of $n_0(t)$ arize. The appearance of these casps means that the motion
has passed to another phase region, as has been discussed in the analysis of 
the previous section.

After crossing the critical line (37), the dynamics of the fractional 
populations remains qualitatively unchanged. Increasing the detuning slightly
changes the oscillation period and smoothes the shape of the oscillation curves,
as is shown in figure 3.

In conclusion, we have presented a theory for the resonant excitation of 
{\it topological coherent modes} of trapped Bose atoms. These modes are the 
stationary solutions to the nonlinear Schr\"odinger equation, which is also
sometimes called the Gross-Pitaevskii equation. This equation provides 
a correct description of trapped atoms at low temperatures [1--3,35]. The 
principally important part of the paper is the demonstration of the occurrence 
of {\it critical dynamic phenomena} in the process of exciting coherent  
modes and a thorough elucidation of the origin of these phenomena.

\vskip 5mm

{\bf Acknowledgements}

\vskip 3mm

One of the authors (V.I.Y.) is very grateful for many useful discussions to
V.K. Melnikov and for important remarks to L.P. Pitaevskii.

\newpage

\begin{center}
{\bf Figure Captions}
\end{center}

\vskip 2cm

{\bf Figure 1}. The fractional populations of the ground coherent mode 
$n_0(t)$ (dashed line) and of the excited topological coherent mode $n_k(t)$ 
(solid line) as functions of time, measured in units of $\al^{-1}$, for the
fixed pumping parameter $b=0.5$ and the negative detuning parameter below
the critical line: (a) $\ep=-10^{-1}$; (b) $\ep=-10^{-2}$; (c) $\ep=-10^{-4}$.

\vskip 1cm

{\bf Figure 2}. Dramatic changes in the dynamics of the fractional populations 
of the ground coherent mode (dashed line) and excited mode (solid line) when
crossing the critical line on the parametric manifold for fixed $b=0.5$ and 
varying detuning: (a) $\ep=0$; (b) $\ep=10^{-9}$.

\vskip 1cm

{\bf Figure 3}. Qualitatively different temporal behaviour of the fractional 
populations of the ground mode (dashed line) and excited mode (solid line)
after crossing the parametric critical line, with fixed $b=0.5$ and varying
detuning: (a) $\ep=10^{-4}$; (b) $\ep=10^{-2}$; (c) $\ep=10^{-1}$.

\end{document}